\documentclass[twoside,twocolumn,9pt]{article}
\usepackage{extsizes}
\usepackage[super,sort&compress,comma]{natbib} 
\usepackage[version=3]{mhchem}
\usepackage[left=1.5cm, right=1.5cm, top=1.785cm, bottom=2.0cm]{geometry}
\usepackage{balance}
\usepackage{mathptmx}
\usepackage{sectsty}
\usepackage{graphicx} 
\usepackage{lastpage}
\usepackage[format=plain,justification=justified,singlelinecheck=false,font={stretch=1.125,small,sf},labelfont=bf,labelsep=space]{caption}
\usepackage{float}
\usepackage{fancyhdr}
\usepackage{fnpos}
\usepackage[english]{babel}
\addto{\captionsenglish}{%
  
}
\usepackage{array}
\usepackage{droidsans}
\usepackage{charter}
\usepackage[T1]{fontenc}
\usepackage[usenames,dvipsnames]{xcolor}
\usepackage{setspace}
\usepackage[compact]{titlesec}
\usepackage{hyperref} 
\hypersetup{hidelinks} 

\usepackage{epstopdf}

\definecolor{cream}{RGB}{222,217,201}

\usepackage{url}

\usepackage{siunitx}

\begin{document}

\pagestyle{fancy}
\thispagestyle{plain}
\fancypagestyle{plain}{
\renewcommand{\headrulewidth}{0pt}
}

\makeFNbottom
\makeatletter
\renewcommand\LARGE{\@setfontsize\LARGE{15pt}{17}}
\renewcommand\Large{\@setfontsize\Large{12pt}{14}}
\renewcommand\large{\@setfontsize\large{10pt}{12}}
\renewcommand\footnotesize{\@setfontsize\footnotesize{7pt}{10}}
\makeatother

\renewcommand{\thefootnote}{\fnsymbol{footnote}}
\renewcommand\footnoterule{\vspace*{1pt}%
\color{cream}\hrule width 3.5in height 0.4pt \color{black}\vspace*{5pt}} 
\setcounter{secnumdepth}{5}

\makeatletter 
\renewcommand\@biblabel[1]{#1}            
\renewcommand\@makefntext[1]%
{\noindent\makebox[0pt][r]{\@thefnmark\,}#1}
\makeatother 
\renewcommand{\figurename}{\small{Fig.}~}
\sectionfont{\sffamily\Large}
\subsectionfont{\normalsize}
\subsubsectionfont{\bf}
\setstretch{1.125} 
\setlength{\skip\footins}{0.8cm}
\setlength{\footnotesep}{0.25cm}
\setlength{\jot}{10pt}
\titlespacing*{\section}{0pt}{4pt}{4pt}
\titlespacing*{\subsection}{0pt}{15pt}{1pt}

\fancyfoot{}
\fancyfoot[LO,RE]{\vspace{-7.1pt}\includegraphics[height=9pt]{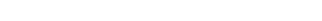}}
\fancyfoot[CO]{\vspace{-7.1pt}\hspace{13.2cm}\includegraphics{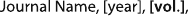}}
\fancyfoot[CE]{\vspace{-7.2pt}\hspace{-14.2cm}\includegraphics{head_foot/RF}}
\fancyfoot[RO]{\footnotesize{\sffamily{1--\pageref{LastPage} ~\textbar  \hspace{2pt}\thepage}}}
\fancyfoot[LE]{\footnotesize{\sffamily{\thepage~\textbar\hspace{3.45cm} 1--\pageref{LastPage}}}}
\fancyhead{}
\renewcommand{\headrulewidth}{0pt} 
\renewcommand{\footrulewidth}{0pt}
\setlength{\arrayrulewidth}{1pt}
\setlength{\columnsep}{6.5mm}
\setlength\bibsep{1pt}

\makeatletter 
\newlength{\figrulesep} 
\setlength{\figrulesep}{0.5\textfloatsep} 

\newcommand{\topfigrule}{\vspace*{-1pt}%
\noindent{\color{cream}\rule[-\figrulesep]{\columnwidth}{1.5pt}} }

\newcommand{\botfigrule}{\vspace*{-2pt}%
\noindent{\color{cream}\rule[\figrulesep]{\columnwidth}{1.5pt}} }

\newcommand{\dblfigrule}{\vspace*{-1pt}%
\noindent{\color{cream}\rule[-\figrulesep]{\textwidth}{1.5pt}} }

\makeatother

\twocolumn[
  \begin{@twocolumnfalse}
{\includegraphics[height=30pt]{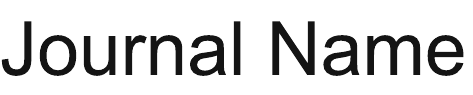}\hfill\raisebox{0pt}[0pt][0pt]{\includegraphics[height=55pt]{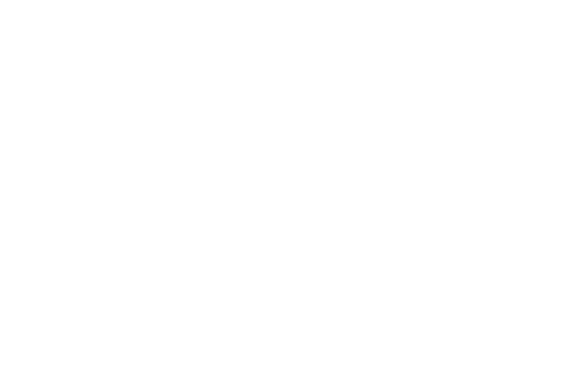}}\\[1ex]
\includegraphics[width=18.5cm]{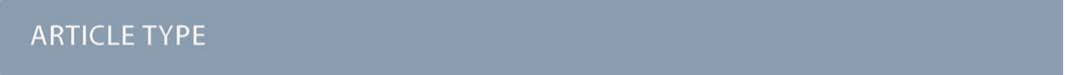}}\par
\vspace{1em}
\sffamily
\begin{tabular}{m{4.5cm} p{13.5cm} }

\includegraphics{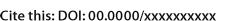} & \noindent\LARGE{\textbf{Mixing with viscoelastic waves at low Reynolds numbers$^\dag$}} \\
\vspace{0.3cm} & \vspace{0.3cm} \\

 & \noindent\large{Enrico Turato,\textit{$^{ab}$} Christelle N. Prinz,\textit{$^{abc}$} Jason P. Beech\textit{$^{ab}$} and Jonas.O Tegenfeldt\textit{$^{ab}$}$^{\ast}$} \\

\includegraphics{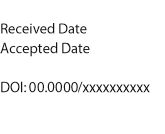} & \noindent\normalsize{Mixing at the microfluidic scale is challenging due to the low Reynolds numbers and often high P\'eclet numbers. Without turbulence, mixing relies solely on diffusion, resulting in slow and inefficient mixing. We demonstrate enhanced mixing in a simple Y-shaped microfluidic channel using viscoelastic turbulence in fluids containing macromolecules, such as DNA and polyethylene oxide. We investigated mixing at two distinct scales: the mixing of small molecules and the mixing of polymers. We show how the viscoelastic fluctuations fold the solvent, resulting in enhanced reaction rate between two reagents. We also show how the viscoelastic turbulence enhances the mixing of the macromolecules. We discuss optimization strategies taking into account mixing efficiency, mixing time, mixing length and energy efficiency. Viscoelastic turbulence unlocks rapid mixing in microfluidic channels where conventional turbulence cannot operate, offering a versatile platform for applications ranging from chemical synthesis to biomedical assays.
} \\

\end{tabular}

 \end{@twocolumnfalse} \vspace{0.6cm}

  ]

\renewcommand*\rmdefault{bch}\normalfont\upshape
\rmfamily
\section*{}
\vspace{-1cm}



\footnotetext{\dag~Supplementary Information available: [details of any supplementary information available should be included here]. See DOI: 00.0000/00000000.\\$^a$Division of Solid State Physics, Department of Physics and NanoLund, Lund University, P.O. Box 118, 22100 Lund, Sweden\\
$^{b} $SciLife Lab, Lund University, P.O. Box 118, 22100 Lund, Sweden
\small$^\ast$Corresponding author. Email: jonas.tegenfeldt@fysik.lu.se
}

\newcommand{\pcc}{\,cm$^{-3}$}	

\section*{Introduction}
\emph{Mixing} is of central importance for numerous applications, even at the microscale \cite{mixing_review_Karnik2015}. At a fundamental level, it relies on diffusion \cite{ taylor1953dispersion, taylor1954conditions, ottino_wiggins}, and the diffusion time depends on the diffusion length quadratically. Therefore, to enhance the mixing rate, a primary strategy is to reduce the diffusion length. \cite{mixing_by_folding_maths_theory}
At the macroscopic scale, inertial turbulence \cite{goldstein1938modern} folds the liquids efficiently and constitutes the main phenomenon responsible for mixing. \cite{dimotakis2005turbulent}
At the microscopic scale, where microfluidics operates, small length scales and flow speeds usually cause viscous effects to dominate over inertial contributions. \cite{brody1996biotechnology}
In this low-Reynolds number regime, laminar flow dominates over inertial turbulence, leaving diffusion as the sole contributor to the mixing, which is very slow.\cite{reynolds1883xxix} To address this limitation, asymmetrically shaped grooves in the bottom part of a fluidic channel can be used to guide the flow into a chaotic pattern that in turn promotes the mixing of the liquids of interest. \cite{chaotic_mixer_stroock2002} Alternatively, 2-layer crossing channel devices can be used to enhance mixing at low-Reynolds numbers \cite{3Darchitecture_mixer}. While effective at enhancing the mixing of two solutions, these methods  involve a 2-layer fabrication process, which can be cumbersome.

Another solution is to add macromolecules to the solution. These polymers take part in an energy exchange with the solvent and this leads to viscoelastic turbulence, where small-scale disruptions cascade into large-scale instabilities and vortices. \cite{elastic_turbulence_Groisman2000, Groisman2001, exp_contr_3} The use of viscoelastic instabilities \cite{quake} for mixing purposes has attracted increasing attention in recent years. \cite{review1, review2,browne2023harnessing, khalilian2025numerical} Viscoelastic instabilities can be achieved by adding small amounts of polymers to a newtonian fluid. For instance, adding small amounts of polyacrylamide in two solutions flowing side by side in a porous medium greatly enhanced their mixing. \cite{datta_mixing2024}

In microfluidic devices, viscoelasticity-driven enhanced mixing has been shown through relatively complex channel geometry modifications, such as long serpentine microchannels, channels with abrupt contractions and expansions, and active components in the device. \cite{topological_mixing, 3Darchitecture_mixer, enhancement_methods, geom_variations_mixers, review_Datta2022, kawale2017, HawardShenReview2021, shnapp2022nonmodal, Groisman2001, exp_contr_3} . The serpentine methods requires very long channels.

Using simple microchannels, we have previously shown that DNA solutions, flowing through pillar arrays at low Reynolds numbers and high Deborah numbers, form viscoelastic instabilities (also called waves) on multiple length scales \cite{strom_waves2023} that depend greatly on the geometry of the array \cite{oskarGeometry2024} and the shapes of the pillars. \cite{beech2023symmetry}
We demonstrated a proof of principle of mixing of parallel flows in pillar arrays in microchannels and showed how a careful selection of the geometry of the array can be used to either enhance or suppress mixing. \cite{beech2023symmetry}

In this work, we show qualitatively and quantitatively how these phenomena can be leveraged for a controlled enhancement of mixing in a Y-shaped microfluidic device in which two fluids streams are flown side by side. As an alternative to DNA, we use polyethylene oxide (PEO) which is highly bio compatible, inert and low cost, and has been shown before to render solutions viscoelastic. \cite{rodd2005inertio, rodd2007role, li2011non, kenney2013large, shi2015mechanisms, shi2016growth} It is therefore better suited than DNA for scaling up the processes and for a broad range of applications. We use fluorescence microscopy to visualise the viscoelastic waves, quantify the mixed fraction and mixing rate. We demonstrate that both solvent and macromolecules undergo enhanced mixing and that this method is more energy-efficient, compared to diffusion-driven mixing.

\section*{Results}

\begin{figure}[!htpb]
\centering
\hspace*
{-0.24cm}\includegraphics[width=0.5\textwidth]{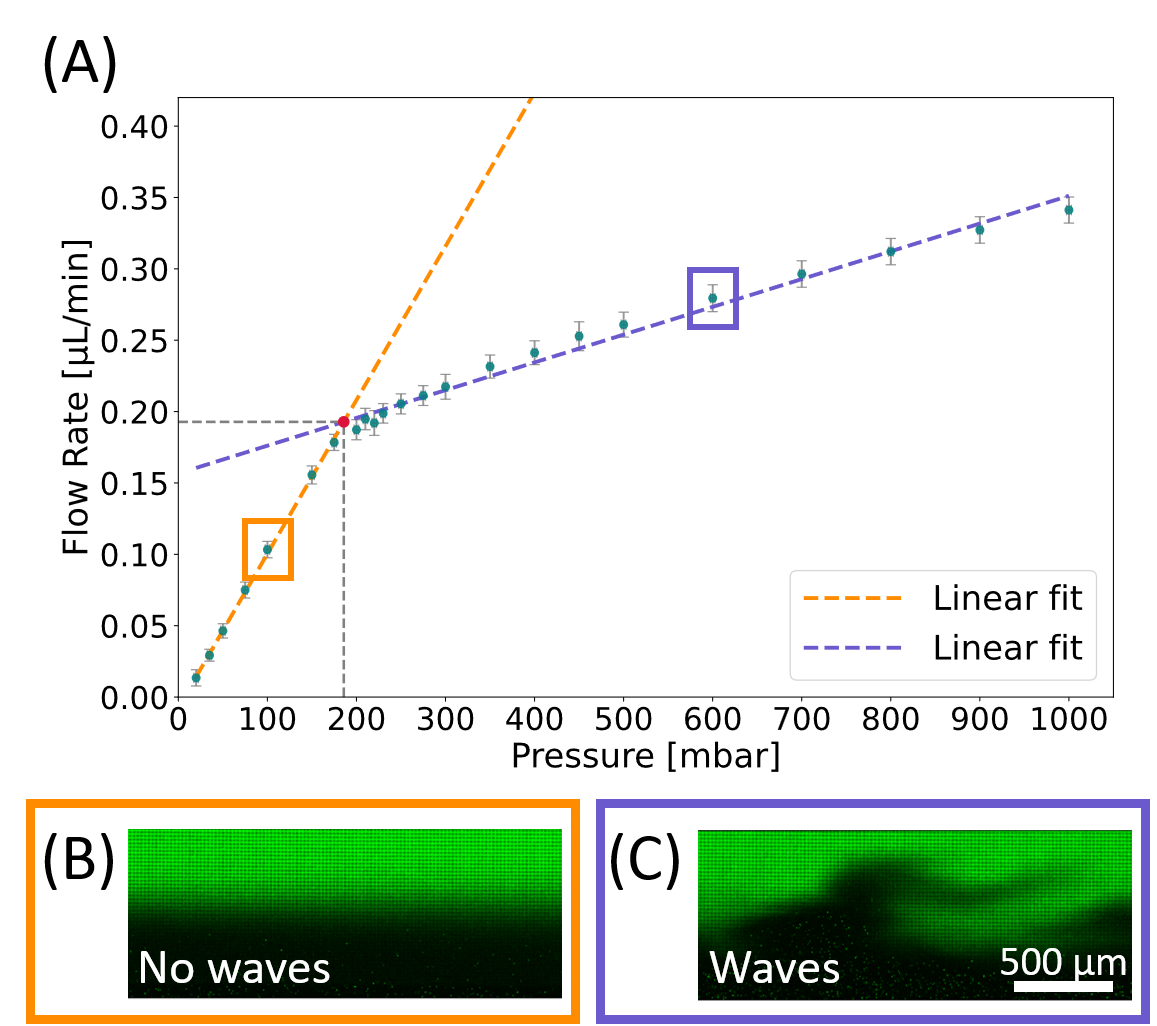}
\caption{\textbf{Mixing of two fluids, one containing Fluorescein.} Both fluids are 0.2\% w/v PEO in water, one has added Fluorescein for visualization. (\textbf{A})  Measured flow rate as a function of applied pressure, showing two linear regimes, with a transition at P $\approx$ 185 mbar. (\textbf{B},\textbf{C})   Fluorescence microscopy images of the microchannel for an applied pressure P\,=\,100 mbar (\textbf{B}), resulting in a laminar flow with purely diffusive mixing, and P\,=\,600 mbar (\textbf{C}), clearly showing the presence of waves with a viscoelasticity-enhanced mixing.} 
\label{fig:fluorescein} 
\end{figure}

To visualize the mixing at the interface of the co-flowing PEO solutions, we added fluorescein to one of them and imaged the flows with fluorescence microscopy. In parallel, we measured  the total flow rate as a function of applied pressure (Fig.~\ref{fig:fluorescein}). Two linear regimes were observed for the flow rate, with a transition at P $\approx$ \qty{185}{\milli\bar} corresponding to a flow rate of \qty{0.2}{\micro\liter} (Fig.~\ref{fig:fluorescein}\textbf{A}).

Fluorescence images of the channel confirmed that the flow is laminar and the mixing therefore purely diffusive below that pressure and enhanced by viscoelastic turbulence above, (Fig.~\ref{fig:fluorescein}(\textbf{B-C})). In all our experiments the Reynolds number is below one, precluding any inertial contribution to the viscoelastic turbulence. For estimated values of Reynolds numbers and other experimental parameters, refer to Table S1 in the Supplementary Material. We would like to point out that in the following, the term \emph{mixing} may refer to a \emph{dilution} of one liquid species into another species (for the fluorescein case and the DNA overlap case), or to a \emph{reaction} between two species combined with a \emph{dilution} of the reactant and the product.


\begin{figure}[!htpb]
\centering
\includegraphics[width=0.5\textwidth]{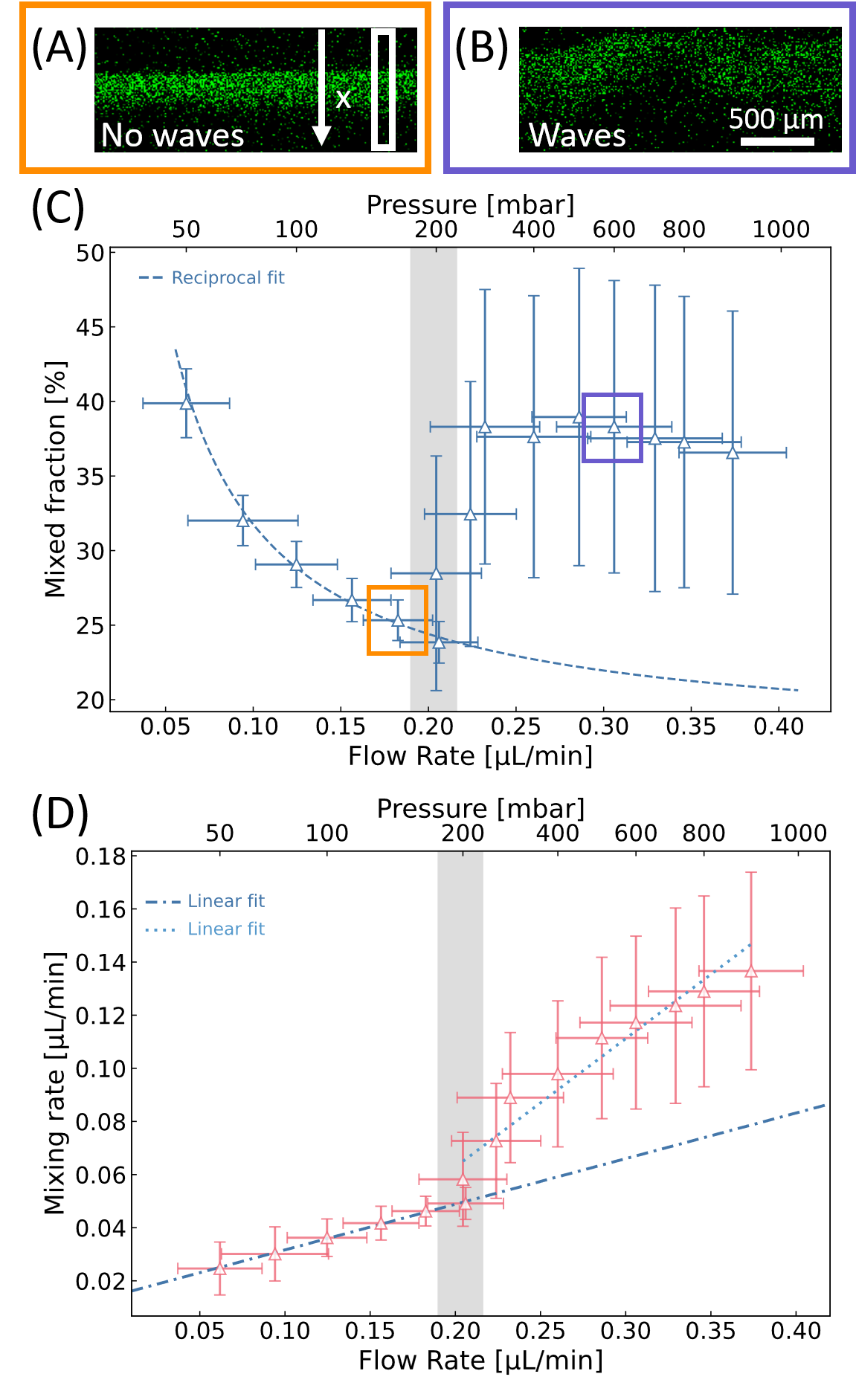}
 \caption{\textbf{Quantifying mixing via fluorescence generation from reactant solutions in the device.} Fluo-3 was added to the 0.2\% w/v PEO aqueous solution and loaded in one inlet. $\text{Ca}^{2+}$ was added to the 0.2\% w/v PEO aqueous solution and loaded in the other inlet. The gray area shows the wave onset, see Fig.~\ref{fig:fluorescein}. (\textbf{A}) Fluorescence microscopy image of the Fluo-3 fluorescence at an applied pressure P = 150mbar, when the flow is laminar. (\textbf{B}) Fluorescence microscopy image of the Fluo-3 fluorescence at an applied pressure P = 600 mbar, when waves are present and the flow is turbulent. (\textbf{C}) Mixed fraction of the Fluo-3 and $\text{Ca}^{2+}$ solutions as a function of flow rate (mean value $\pm$SD). The dashed line is a reciprocal fit corresponding to a purely diffusional mixing (\textbf{D}) Mixing rate of the two solutions as a function of flow rate (mean value $\pm$SD). The dashed lines are linear fits of the rates with and without waves.} 
\label{fig:3_reaction} 
\end{figure}

To quantify the mixing, we repeated the experiments using the calcium indicator Fluo-3 in one inlet and $\text{Ca}^{2+}$ in the other inlet \cite{ismagilov2000}. Fluo-3 comprises a fluorescein moiety and a chelating moiety and binds $\text{Ca}^{2+}$ in a 1:1 stoichiometry ($K_d =0.39\mu M$). Fluorescence is enhanced 40 -- 100 times upon binding to $\text{Ca}^{2+}$. Therefore, although less intense than in the case of fluorescein, the Fluo-3 fluorescence is a more accurate indicator of mixing (Fig.~\ref{fig:3_reaction}). As in the case of fluorescein, the flow rate as a function of applied pressure follows two linear regimes, with a transition at P $\approx$ \qty{170}{\milli\bar} (Fig. S2\textbf{B}). The mixing is purely diffusional at low flow rates, (Fig.~\ref{fig:3_reaction}\textbf{A}) and turbulent at high flow rates (Fig.~\ref{fig:3_reaction}\textbf{B}). 

\clearpage
 We calculate the mixed fraction by averaging the normalized pixel values in the region of interest (ROI, white frame in Fig.~\ref{fig:3_reaction}\textbf{A}), positioned at the end of the microfluidic channel, at $\approx$\,6.3 mm from the point where the two fluids first meet. With an excess of $Ca^{2+}$, the fluorescence intensity in each pixel is proportional to the concentration of Fluo-3. We therefore assume that the maximum pixel value where the two solutions meet corresponds to the inlet concentration of the dye. For full mixing, the average pixel value for the whole channel width thus equals half the maximum value.

\begin{equation}
    \text{Mixed Fraction} = 2 \times \frac{1}{N}\sum_{k \in \text{ROI}} I_{reaction}^{k}\, ,
\label{eq:mixingfraction_react}
\end{equation}

and

\begin{equation}
    \text{Mixing Rate} =  \text{Mixed Fraction} \times Q \, ,
\label{eq:mixingrate_react}
\end{equation}

\noindent where $N$ is the number of pixels in the ROI. $I_{reaction}^{k}$ is the normalized intensity of the fluorescence signal of the Fluo-3--Ca$^{2+}$ complex in pixel $k \in \text{ROI}$, and $Q$ is the flow rate. Plots of the resulting averages over all frames is finally shown in Fig.~\ref{fig:3_reaction}\textbf{C} and \textbf{D}. At flow rates when the flow is laminar, the mixed fraction and mixing rates follow the expected values for a diffusion-dominated mixing (reciprocal fit in Fig.~\ref{fig:3_reaction}\textbf{C} and lineal fit in Fig.~\ref{fig:3_reaction}\textbf{D}). As the flow rate increases and the flow becomes turbulent, there is a steep increase in the mixed fraction compared to a purely diffusional mixing that would take place at the same flow rate (dashed line). The mixed fraction reaches a plateau at even higher flow rates (Fig.~\ref{fig:3_reaction}\textbf{C}). The mixing rate increases linearly with the flow rate, but the increase is steeper when waves are present Fig.~\ref{fig:3_reaction}\textbf{D}.

\begin{figure}[!htpb]
\centering
\includegraphics[width=0.5\textwidth]{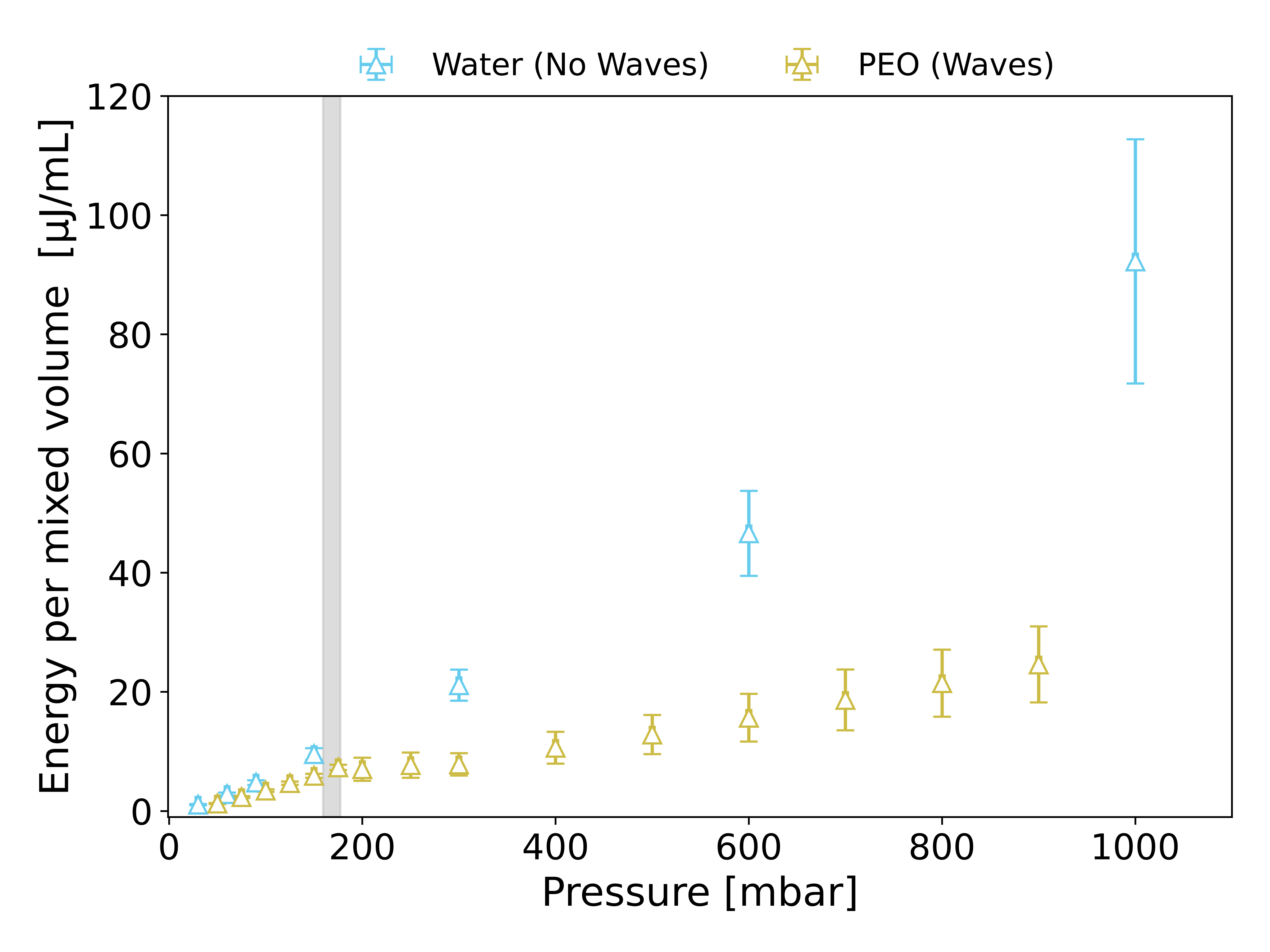}
\caption{\textbf{Energy cost per mixed volume of mixing a Fluo-3 solution and a $\text{Ca}^{2+}$ solution in the device.} The energy efficiency is compared for two fluids: Newtonian fluid, water (blue symbols) and a viscoelastic fluid, 0.2\% w/v PEO aqueous solution (yellow symbols). The gray region indicates the estimated onset of elastic viscoelastic instabilities (waves).}
\label{fig:energy} 
\end{figure}

Next, we investigated the energy efficiency of the viscoelasticity driven mixing of Fluo-3 and $\text{Ca}^{2+}$ solutions. Based on the simple fact that the unit of pressure can be written in terms of energy per volume, $1\mathrm{Pa}=1\tfrac{\mathrm{J}}{\mathrm{m^3}}$, we calculated the energy per mixed volume for each applied pressure tested in our device and compared it with the case of using water without PEO as solvent (Fig.~\ref{fig:energy}).

At low pressures, when no waves are present in the PEO solution, there are no significant differences in energy costs per mixed volume between both fluids. When waves are present in the PEO solution (P $\ge$ 200 mbar), the energy cost of mixing is substantially ($\approx3\times$) lower for the viscoelastic fluid, compared to when water is used. Combined with the results of Fig.~\ref{fig:3_reaction}, this shows that viscoelasticity-driven mixing not only yields a greater amount of mixed solution per unit time, but is also more energy-efficient.

To this point, we have demonstrated that the mixing of small molecules, such as Fluo-3 and $\text{Ca}^{2+}$ in a 0.2\% (w/v) aqueous PEO solution is significantly enhanced and becomes more efficient in the presence of viscoelastic instabilities. We further sought to determine whether a similar enhancement applies to the polymers themselves. Because PEO cannot be directly fluorescently labeled, we instead loaded two differently colored macromolecular DNA solutions (red and green) in buffer (no PEO) in each inlet and quantified polymer mixing.

\begin{figure}[!htpb]
\includegraphics[width=0.5\textwidth]{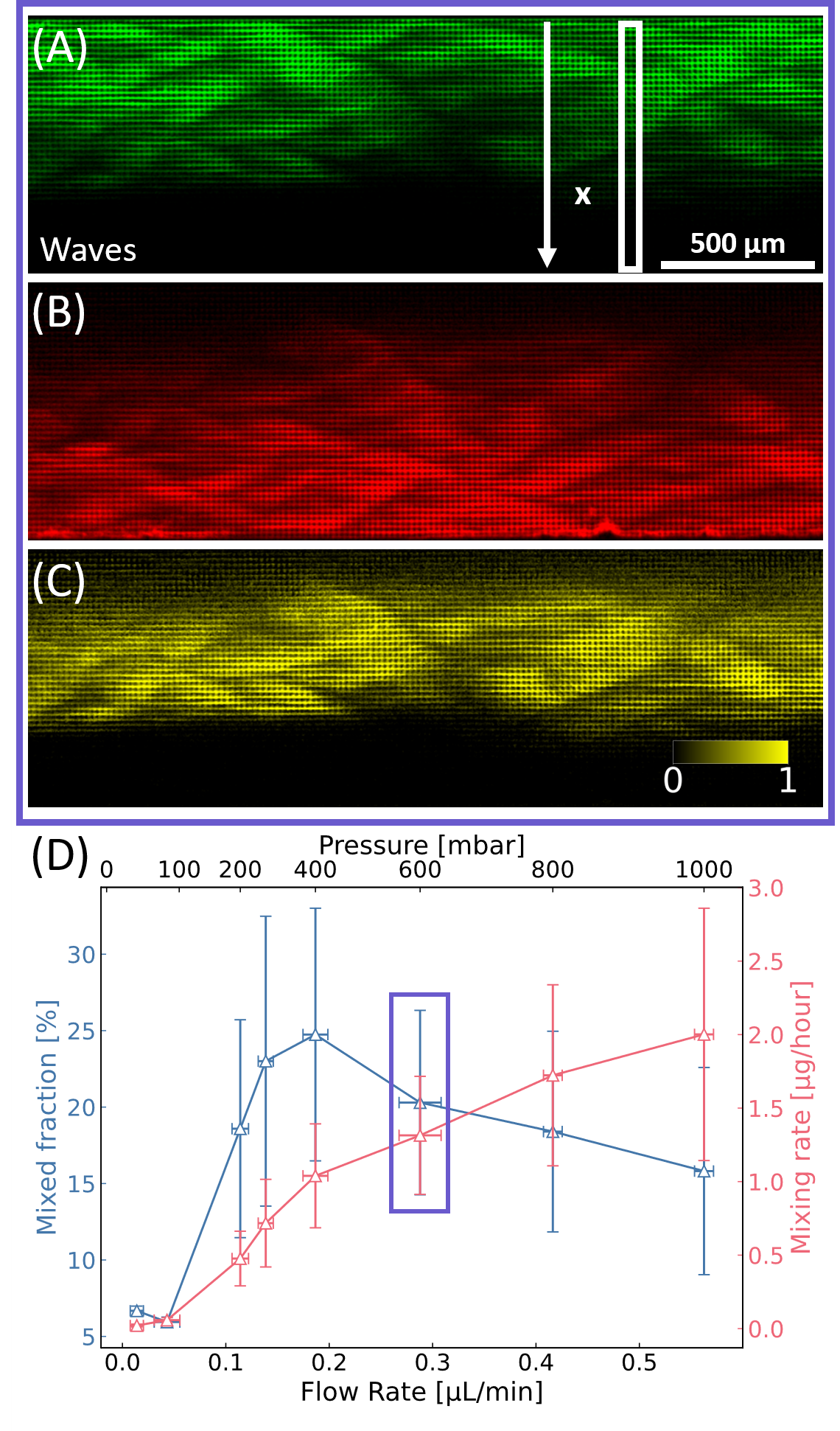}
\centering
\caption{\textbf{Mixing performance of DNA macromolecule solutions in the device.} (\textbf{A}) Fluorescence microscopy image of the green DNA, showing viscoelastic waves. (\textbf{B}) Fluorescence micropscopy image of the red DNA. (\textbf{C}) Overlap of the two colors, representing the mixing of the 2 DNA, with color scale from 0 to 1.(\textbf{D}) mixed fraction, (mean value $\pm$SD) and Mixing rate (mean value $\pm$SD). Spline curves are included to guide the eye. }
\label{fig:DNA} 
\end{figure}

We calculate the mixed fraction of the two DNA solutions by first averaging the normalized pixel values along the channel direction in the ROI (white frame in Fig.~\ref{fig:DNA}\textbf{A}), positioned at the end of the microfluidic channel, at $\approx$\,6.3 mm from the point where the two fluids first meet. We then take an average of the minimum of the red and green intensities in each pixel position. Full mixing results in the concentration of each DNA solution to be half of its value as it enters the device. With the intensity proportional to the concentration of the DNA, we assume that full mixing corresponds to pixel values half of the maximum values. 


\begin{equation}
    \text{Mixed Fraction} = 2 \times \frac{1}{N}\sum_{i}^{N} min(\frac{1}{M}\sum_{j}^{M} I_{red}^{i, j}\, , \frac{1}{M}\sum_{j}^{M} I_{green}^{i, j}),
\label{eq:DNA}
\end{equation}

\noindent where $I_{red}^{i,j}$ is the intensity of the red fluorescence in pixel $(i, j)$, and $I_{green}^{i,j}$  is the intensity of the green fluorescence. The minimum value is selected after averaging along the channel, $y$, direction in the ROI.
The Mixed fraction and Mixing Rate are finally averaged over all frames and shown in (Fig.~\ref{fig:DNA}).

 The mixing of both DNA solutions is greatly enhanced compared to when no turbulence is present. The mixed fraction is maximal at a flow rate of \qty{0.2}{\micro\liter / \minute} (\qty{400}{\milli\bar} applied pressure), however, the mixing rate continues to increase with flow rate beyond this value.

\section*{Discussion}

We report controlled enhancement of mixing of both small molecules and polymers in a Y-shaped microfluidic channel containing a pillar array. The system exhibits two distinct flow regimes: laminar flow at low flow rates and a viscoelastic instability at higher flow rates characterized by the onset of viscoelastic waves. The transition to this wave-dominated regime leads to a marked increase in both the extent of mixing and the mixing rate. These results reinforce the concept that viscoelastic flow instabilities can be harnessed as an efficient route to overcome diffusion-limited transport at low Reynolds numbers.

A key practical advantage of the present design is its compact footprint. The total mixing length (8 mm) is substantially shorter than that of many passive micromixers based on geometrical perturbations such as staggered herringbone structures \cite{chaotic_mixer_stroock2002} or earlier serpentine-design viscoelastic mixing platforms. When operated above the instability threshold, the device achieves mixed fractions of 30–40\% in less than \qty{4}{\second}, Fig. S3 and S4. Although complete homogenization is not reached within the device length studied here, the spatial distribution of concentration reveals that the mixed fluid is preferentially located near the channel center. This feature suggests straightforward strategies for downstream integration, such as multi-outlet architectures that separate well-mixed from partially mixed streams. Guiding the latter to the inlet of a subsequent device could, in principle, enable near-complete utilization of reagents without increasing device length.

Beyond mixing performance alone, our results highlight the advantages of viscoelastic-driven mixing from the perspective of energy conservation. For comparable levels of mixing, the viscoelastic flows require less energy input than Newtonian flows relying solely on diffusion. Energy efficiency is a critical consideration for portable or autonomous microfluidic systems with limited power budgets \cite{wearables_energylimitation_perspective2022, wearables_limitations2023}, as well as for large-scale parallelization where operating costs scale with pressure drop. \cite{Kitamori_parallel2025} The present findings therefore position viscoelastic instabilities not only as a transport-enhancing mechanism but also as a lever for reducing the energetic cost per processed volume.


The data further show that optimization of such systems requires balancing competing objectives. From an energy perspective, there is a penalty for high applied pressures. The mixed fraction saturates above the threshold flow rate for the solvent mixing (Fig.~\ref{fig:3_reaction}) and it has a peak for the macromolecular mixing (Fig.~\ref{fig:DNA}). However, the mixing rates increase with increased pressures for both cases. Similarly, as a function of position in the device, the mixed fraction saturates approximately half way along the mixing array for the solvent mixing (Fluo-3 and $\text{Ca}^{2+}$), Fig. S3, and it increases monotonously for the macromolecular mixing, Fig. S4.


Keeping in mind that the mixing enhancement takes place at the threshold flow rate, a shorter and deeper device would lower the required pressure to reach the threshold flow rate and thus the energy per processed volume. In order to scale up while keeping energy usage to a minimum, each mixer should be made shorter and deeper and be connected in parallel rather than applying an increased applied pressure to the whole device.

Taken together, these results demonstrate that pillar-induced viscoelastic instabilities provide a compact and energetically efficient strategy for microfluidic mixing of both small molecules and polymers. By clarifying the relationships among flow rate, position, species type, and energy cost, this work establishes design principles for next-generation viscoelastic micromixers and supports their integration into scalable, resource-efficient microfluidic platforms.

\section*{Materials and Methods}
\subsection*{Reagents}

Aqueous solution containing 0.2\% [wt/vol] linear polyethylene oxide (PEO) MW $\approx$ 8 MDa (Sigma-Aldrich, St. Louis, MO, USA) was used as the viscoelastic solution. To visualize the waves, Fluorescein sodium salt (Sigma-Aldrich, Merck KGaA, Darmstadt, Germany) was used at a concentration of 0.1 M MilliQ water. As a simple chemical reaction, we exploited the interaction between Fluo-3 and $\text{Ca}^{2+}$. Fluo-3 is a pentapotassium salt (Sigma-Aldrich, Merck KGaA, Darmstadt, Germany), which is a commercially available non-fluorescent compound, that becomes fluorescent when reacting with calcium ions. We used an aqueous solution of calcium chloride, $\text{CaCl}_2$, (Sigma-Aldrich, Merck KGaA, Darmstadt, Germany). One inlet was filled with 4 mM CaCl$_2$ in 0.2 \% [w/v] PEO in MilliQ water. The other inlet was filled with $10 \mu M$ Fluo-3 in a solution of 0.2 \% [w/v] PEO in 0.33$\times$ Tris-EDTA \cite{ismagilov2000}, corresponding to 3.3 mM Tris and 0.33mM EDTA. The chelating agent EDTA in the Fluo-3 solution ensures that any Ca$^{2+}$ ions are removed to minimize any fluorescence from the Fluo-3 that is not mixed with the $\text{CaCl}_2$ solution. Note that a concentration of the EDTA is selected to be an order of magnitude less than that of the Ca$^{2+}$.

For the DNA solution,  we used \qty{375}{\nano\gram / \micro\liter}  of $\lambda$ phage DNA (New England Biolabs, Ipswich, MA, USA) in a solution of $5\times$ Tris EDTA (TE) buffer. The DNA was stained with two bis-intercalating dyes YOYO-1 (green) or YOYO-3 (red) (Life Technologies, Carlsbad, CA, USA) to create two DNA sub-populations that were subsequently mixed. The staining ratio was DNA bp:dye 50:1. See the Supplementary Information for detailed protocol.

\subsection*{Device fabrication}
\begin{figure}[!htpb]
\centering
\includegraphics[width=0.5\textwidth]{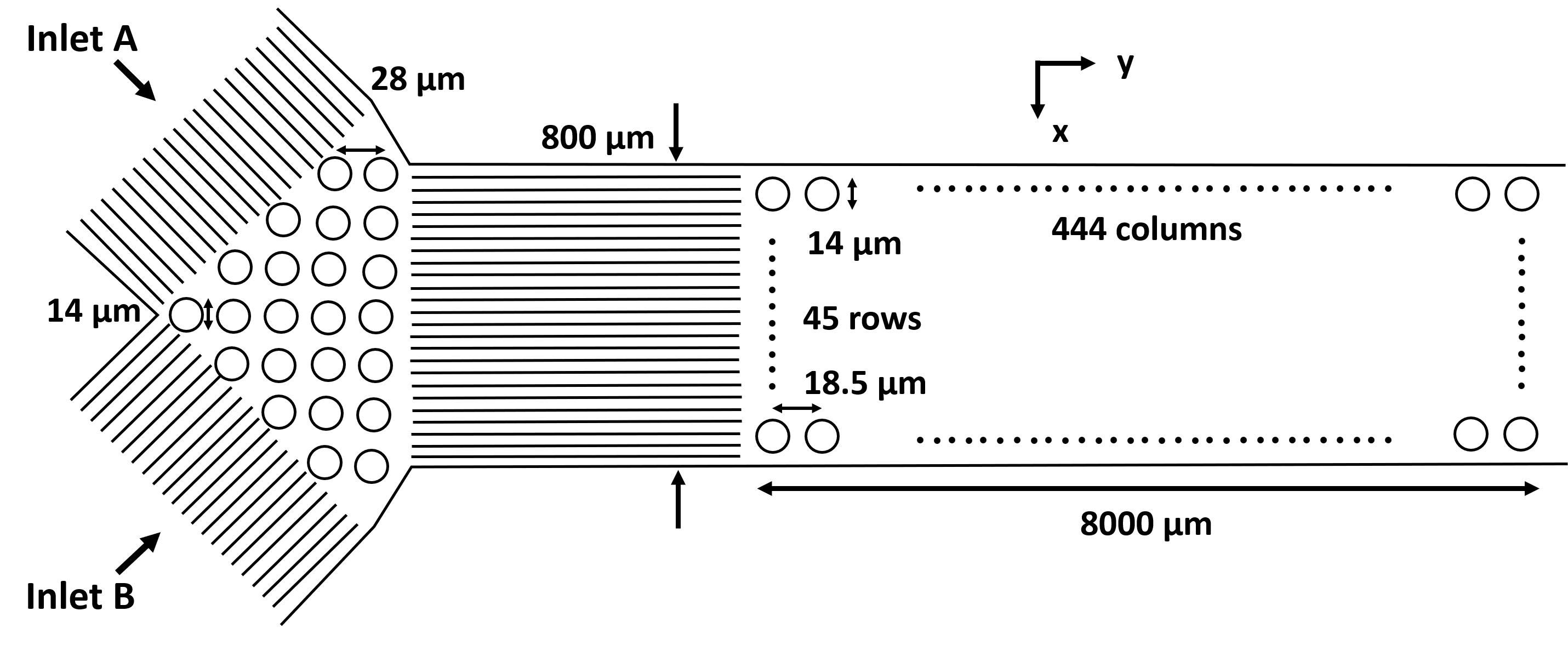}
\caption{\textbf{Schematics of the device (not to scale).} The inlets contain channels and support pillars. The main channel is \qty{800}{\micro\meter} $\times$ \qty{8000}{\micro\meter} $\times$ \qty{11}{\micro\meter} and contains a square array of pillars (\qty{14}{\micro\meter} in diameter, with a pitch of \qty{18.5}{\micro\meter}). The $x$ and $y$ directions are shown in the figure. The $z$ direction points out of the device plane.} 
\label{fig:device} 
\end{figure}

The device design was produced using the L-Edit software (Tanner Research, Monrovia, CA, USA), the master was fabricated with a MLA150 maskless lithography system (Heidelberg Instruments GmbH, Heidelberg, Germany) on a silicon wafer, followed by standard soft lithography procedures \cite{soft_litho} to realize the microfluidic devices via standard peel-off procedures. The device is a Y-shape channel, of width $\approx$ \qty{800}{\micro\meter} and height $\approx$ \qty{11}{\micro\meter}, Fig.~\ref{fig:device}. The channel has parallel microchannels and support pillars close to the inlet and a square array of circular pillars further away in the channel where the viscoelastic fluctuations take place. The pillars are \qty{14}{\micro\meter} in diameter and the gap between pillars is \qty{4.5}{\micro\meter}.

\subsection*{Experimental setup}
The flow inside the microchannels was generated by applying nitrogen gas overpressure, controlled with an MFCS-4C pressure controller (Fluigent, Paris, France) and by the software OxyGEN (v. 1.3.0.0, 2021, Fluigent, Paris, France). The flow rate was measured using a flow sensor (Flow rate platform with flow unit S, Fluigent, Paris, France) that was connected to the outlet tubing.

The data were acquired using a standard Nikon epifluorescence microscope (Eclipse Ti microscope, Nikon Corporation, Tokyo, Japan) with a $2 \times$ objective (Nikon Plan UW, NA 0.06) and a $4 \times$ objective (Nikon Plan Apo $\lambda$, NA~0.2) and with the acquisition software NIS-elements AR (v.5.02.03).  A SOLA Light Engine (6-LCR-SB, Lumencor Inc, Beaverton, OR, USA) light source was used. 

For the experiments with \emph{fluorescein}, we used a FITC filter cube and a sCMOS camera (ORCA-Flash4.0 V2, C11440-22CU, Hamamatsu, Hamamatsu, Japan) with $2048 \times 2048$ pixels,  \qty{6.5}{\micro\meter} $\times$ \qty{6.5}{\micro\meter} pixel size and \qty{13.3}{\milli\meter} $\times$ \qty{13.3}{\milli\meter} sensor size.

For the experiments with \emph{Fluo-3/$\text{Ca}^{2+}$}, we used a FITC filter cube and a EMCCD camera (iXon DU--897, Andor Technology, Belfast, Northern Ireland) with $512 \times 512$ pixels,  \qty{16}{\micro\meter} $\times$ \qty{16}{\micro\meter} pixel size and \qty{8.2}{\milli\meter} $\times$ \qty{8.2}{\milli\meter} sensor size.

For the experiments with \emph{DNA}, we used an Optosplit II (Cairn Research Ltd., Faversham, UK) and a sCMOS camera (ORCA-Flash4.0 V2, C11440-22CU, Hamamatsu, Hamamatsu, Japan) with $2048 \times 2048$ pixels,  \qty{6.5}{\micro\meter} $\times$ \qty{6.5}{\micro\meter} pixel size and \qty{13.3}{\milli\meter}~$\times$~\qty{13.3}{\milli\meter} sensor size.

\subsection*{Video processing and quantification of mixing}
Image analysis was performed to quantify the mixing in our device. The fluorescence image data is first processed to compensate for non-uniform illumination, background fluorescence and external background light. A gaussian blur is then performed to remove the pillars from the images. For two-color experiments, the two views representing the two colors were carefully aligned.

The Mixed Fraction was measured in a region of interest (ROI) that can be placed anywhere along the device. The ROI is a rectangle spanning across the microchannel (800 µm), see white rectangle in Fig.~\ref{fig:3_reaction}, and Fig.~\ref{fig:DNA}. Its width was adjusted for each experiment, in order to minimize fluctuations between frames, while capturing the mixing as a function of the position along the microchannel (128 µm for Fluo-3/$\text{Ca}^{2+}$ experiments and 104 µm for DNA experiments). To decrease noise, the pixel values were averaged along the flow direction within the ROI.

For the chemical reaction, the pixel values were normalized by mapping them to $[0, 1]$ using the global minimum and maximum averaged values in all ROI. In contrast, for the DNA overlap calculations, the pixel values were normalized based on the minimum and maximum averaged values in each frame.

When using Fluo-3 and $\text{Ca}^{2+}$, 1000 frames and 200 frames were analyzed in experiments with and without waves, respectively. For DNA experiments, 250 frames were analyzed. For each ROI position along the channel, the fluorescence intensity profile across the channel was extracted and time-averaged over all frames. The computed values are then mapped to their corresponding positions along the channel length.  Since the flow speed is known, the ROI position can be converted into mixing time, allowing the Mixed Fraction to be expressed as a function of time. Unless otherwise noted, the ROI was positioned at a distance $\approx$ \qty{6.3}{\milli\meter} from the beginning of the array. Analyses made using a ROI located at other positions along the channel are shown in Fig. S3 and S4.

\section*{Conclusions}
The conclusions section should come in this section at the end of the article, before the Author contributions statement and/or Conflicts of interest statement.

\section*{Author contributions}
The author contribution statement has been written according to CRediT (Contributor Roles Taxonomy, see https://credit.niso.org/ for role descriptions): Conceptualization, E.T., C.N.P., J.P.B and J.O.T.; Data Curation, E.T.; Formal Analysis, E.T.; Funding Acquisition, J.O.T.; Investigation, E.T., C.N.P., J.P.B and J.O.T.; Methodology, E.T., C.N.P., J.P.B and J.O.T.; Project Administration, J.O.T.; Resources, J.O.T.; Software, E.T.; Supervision, C.N.P., J.P.B., and J.O.T.; Validation, E.T., C.N.P., J.P.B and J.O.T.; Visualization, E.T., C.N.P., J.P.B and J.O.T.; Writing Original Draft, E.T.; Writing - Reviewing \& Editing, E.T., C.N.P., J.P.B. and J.O.T.

\section*{Conflicts of interest}
There are no conflicts to declare.

\section*{Data availability}
The data that support the findings of this study is openly available in the Harvard Dataverse at https://doi.org/10.7910/DVN/HM0XPF . Supplementary movies are available at https://av.tib.eu/series/2005

\section*{Acknowledgements}
All device processing was conducted within Lund Nano Lab.
We have used AI-assisted tools (ChatGPT) for facilitating the translation of conceptual ideas into Python code during the image processing, data analysis and visualization. This research was funded by the European Union, grant number 634890 (project BeyondSeq/Horizon2020), EuroNanoMed (NanoDiaBac), by the Swedish Research Council, grant number
2016-05739 and NanoLund, grant numbers p20-2019, staff01-2020, s01-2024.



\balance


\bibliography{MIX_main}
\bibliographystyle{rsc}

\end{document}